\documentclass[twocolumn,prl,superscriptaddress]{revtex4}
\usepackage{graphicx}
\usepackage{amssymb,amsmath}
\usepackage{bm}
\usepackage{dcolumn}
\usepackage{subfigure}
\usepackage{float}
\usepackage{url}
\usepackage{color}
\usepackage{txfonts}
\usepackage[driverfallback=dvipdfm]{hyperref}
\hypersetup{pdfpagemode=FullScreen,colorlinks=true,breaklinks,urlcolor=blue,linkcolor=blue,citecolor=blue}
\usepackage{natbib}

\begin{document}

\title{Resonant Driving induced Ferromagnetism in the Fermi Hubbard Model}

\author{Ning Sun}
\affiliation{Institute for Advanced Study, Tsinghua University, Beijing, 100084, China}
\author{Pengfei Zhang}
\affiliation{Institute for Advanced Study, Tsinghua University, Beijing, 100084, China}
\author{Hui Zhai}
\affiliation{Institute for Advanced Study, Tsinghua University, Beijing, 100084, China}
\affiliation{Collaborative Innovation Center of Quantum Matter, Beijing, 100084, China}

\date{\today }

\begin{abstract}

In this letter we consider quantum phases and the phase diagram of a Fermi Hubbard model under periodic driving that has been realized in recent cold atom experiments, in particular, when the driving frequency is resonant with the interaction energy. Due to the resonant driving, the effective Hamiltonian contains a correlated hopping term where the density occupation strongly modifies the hopping strength. Focusing on half filling, in addition to the charge and spin density wave phases, large regions of ferromagnetic phase and phase separation are discovered in the weakly interacting regime. The mechanism of this ferromagnetism is attributed to the correlated hopping because the hopping strength within a ferromagnetic domain is normalized to a larger value than the hopping strength across the domain. Thus, the kinetic energy favors a large ferromagnetic domain and consequently drives the system into a ferromagnetic phase. We note that this is a different mechanism in contrast to the well-known Stoner mechanism for ferromagnetism where the ferromagnetism is driven by interaction energy. 

\end{abstract}

\maketitle

Recently, with the help of quantum gas microscope for fermions \cite{microscope-1,microscope-2,microscope-3,microscope-4,microscope-5,microscope-6}, tremendous experimental progresses have been made on quantum simulation of the Fermi Hubbard model. These progresses include the observation of equilibrium properties such as short-range antiferromagnetic correlations \cite{short_AF-1,short_AF-2,short_AF-3}, hidden antiferromagnetic correlations \cite{Bloch_chain}, incommensurate spin correlations \cite{Bloch_incommensurate_chain}, canted antiferromagnetic correlations \cite{canted} and pairing correlations \cite{pairing} in several different circumstances. In particular, the antiferromagnetic quasi-long-range order has been successfully observed through entropy engineering \cite{long_AF}. These progresses also include the study of non-equilibrium transport behaviors such as the measurement of optical conductivity \cite{conductivity}, and the spin and charge transport behavior in the strongly interacting regime \cite{spin, bad_metal}. 

Studying Fermi Hubbard model with cold atoms also allows us to open up new avenue beyond the traditional condensed matter paradigm. One of such examples is the periodically driven Fermi Hubbard model \cite{th1,th2}. Since the typical parameters of a Hubbard model is the hopping strength $J$ and the on-site interaction $U$, both of which are of the order of electron volt in strongly correlated solid-state materials, it is therefore hard to drive a solid-state material with frequency resonant with any of these two energy scales. However, in cold-atom optical lattice realization of the Fermi Hubbard model, the typical energy scales for these two parameters are both of the order of thousand Hertz, and it is quite easy to drive the optical lattices with such a frequency. When the driving frequency is resonant with the interaction parameter $U$, the driving can strongly modify the Fermi Hubbard model. As observed in a recent experiment from the ETH group, the short-range antiferromagnetic correlation can be reduced, or enhanced, or even switch sign to become ferromagnetic correlation \cite{ETH_driven}. Similar experiment has also been performed by driving the Hubbard model with two-photon Raman transition \cite{UIUC}. Hence, by combining such a resonant driving with the quantum gas microscope, it is very promising to study novel physics induced by periodic driving that cannot be accessed in a static system. The goal of this letter is therefore to predict quantum phases and phase diagram of the resonant driven Fermi Hubbard model that is newly realized in cold atom experiments. 

\begin{figure}[t]
	\centering
	\includegraphics[width=0.7\columnwidth]{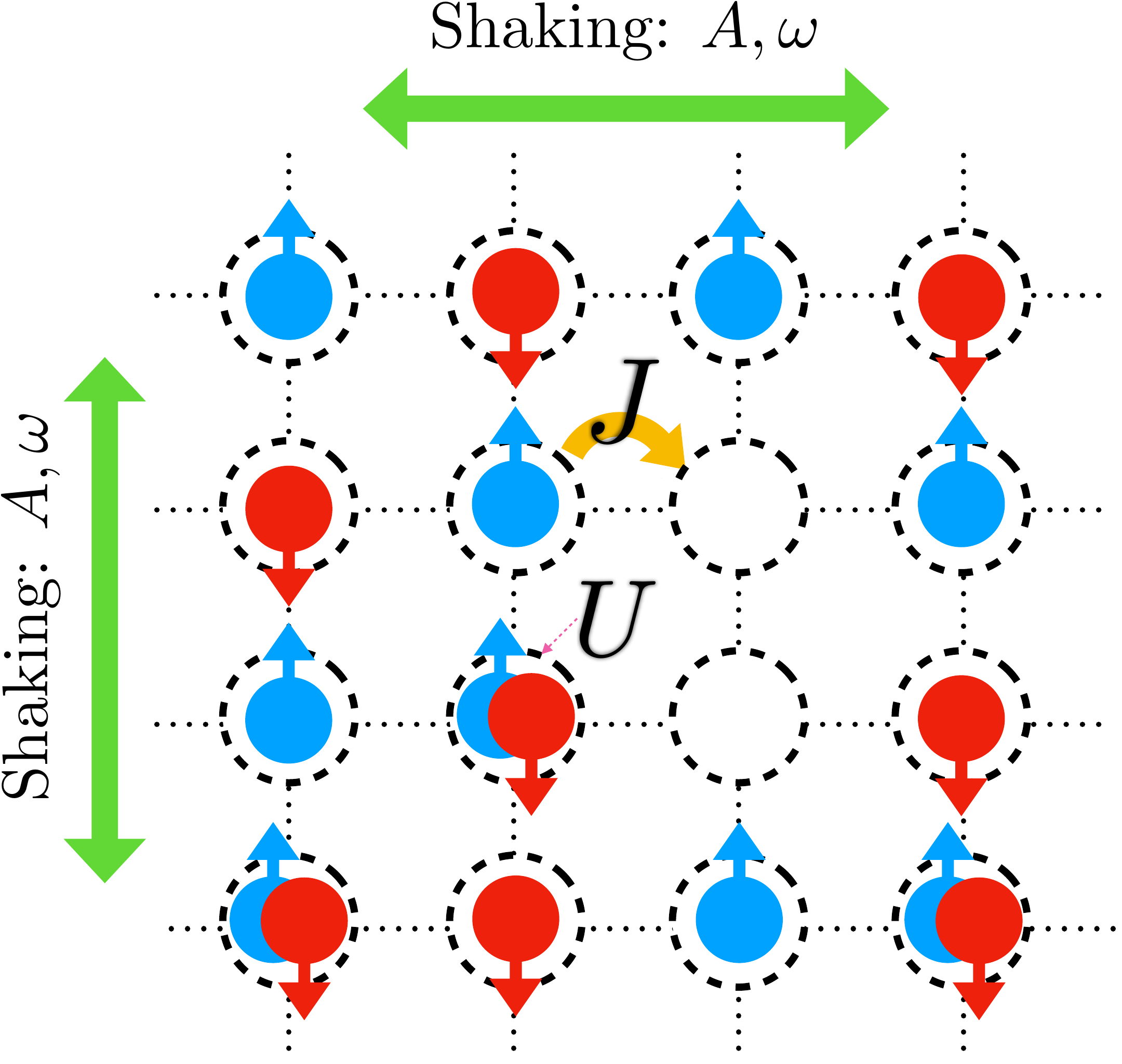}
	\caption{A schematic of the Fermi Hubbard model on a two-dimensional square lattice. $J$ denotes the hopping strength. $U$ is the on-site interaction. The system is time-periodically modulated by shaking the lattice in a sinusoidal way with frequency $\omega$ and amplitude $A$. Arrows indicate the direction of shaking. Balls with different colors and arrows indicate fermions with different spins. }\label{fig:schematic}
\end{figure}

\textit{Model.} We consider a two-dimensional square lattice under similar driving as realized in experiments \cite{ETH_driven,new_exp}. The lattice is periodically modulated along the $\hat{x}$ and $\hat{y}$ directions with a frequency $\omega$ and an amplitude $A$, whose single particle Hamiltonian can be written as
\begin{align}
	\hat{H}_{0}(t) = \frac{{\bf p}^2}{2m} + \hat{V}({\bf x}+A\cos(\omega t),{\bf y}+A\cos(\omega t)) \label{H01}
\end{align}
where $m$ is the mass of an atom. We can now preform a unitary transformation \cite{unitary}
\begin{align}\label{eq:U}
	\hat{U}(t) = \exp(i{\bf p}\cdot{\bf r}_0(t)),
\end{align}
where ${\bf r}_0(t) = -A\cos(\omega t)(1,1)$ are two-dimensional vectors. This unitary transformation transfers position ${\bf r}_0(t)$ into the comoving frame where the lattice becomes static but an extra time-dependent gauge field is introduce. The resulting Hamiltonian is written as 
\begin{align}
	\hat{H}_{0}(t) = \frac{({\bf p}-{\bf A(t)})^2}{2m} + \hat{V}({\bf r}), \label{H02}
\end{align}
with ${\bf A}(t)=m {\bf r}_0(t)$. In principle, the Hamiltonian Eq. \ref{H02} is equivalent to the one Eq. \ref{H01} but it is more convenient for later purpose.

Now we consider a single-band tight-binding model with the nearest neighboring tunneling coefficient $J$ and on-site Hubbard interaction strength $U$. The Hamiltonian in a second quantized form can be written by the Peierls substitution as
\begin{align}\label{eq:H}
	\hat{H}(t) =& -J \sum_{\substack{\langle i,j\rangle \\ \sigma = \uparrow,\downarrow}}e^{i\mathbf{d_{ij}}\cdot \mathbf{A}(t)} \hat{c}_{i}^{\dagger}\hat{c}_{j}+U\sum_{i}\left(\hat{n}_{i\uparrow}-\frac{1}{2}\right)\left(\hat{n}_{i\downarrow}-\frac{1}{2}\right).
\end{align}
where $\hat{c}_{i\sigma}$ ($\hat{c}_{i\sigma}^{\dagger}$) is the fermionic annihilation (creation) operator on site $i$ with spin $\sigma$, $\hat{n}_{i\sigma}$ is the density operator on site $i$ with spin $\sigma$, $\langle\ldots\rangle$ denotes the nearest neighboring sites, and $\mathbf{d_{ij}}=\mathbf{d_{i}}-\mathbf{d_{j}}$ with $\mathbf{d_i}$ being the position of the $i$th lattice site. Throughout this work we focus on the half-filling case and the chemical potential is set to zero.

If the modulation frequency $\omega$ is the largest energy scale of the problem, one can make a high-frequency expansion to obtain an effective time-independent Hamiltonian \cite{highfreqRef-1,highfreqRef-2}. The effective Hamiltonian takes the same form as the normal Hubbard model and the only modification is that the tunneling coefficient is renormalized by the oscillating gauge field as $\tilde{J} = J\mathcal{B}_0(\mathcal{A})$, where we use $\mathcal{B}_l$ to denote the $l$th Bessel function and $\mathcal{A} = mA\omega d$ is the normalized shaking amplitude hereinafter. $d$ is the distance of two Wannier wave packets in the nearest neighboring lattice sites. 

However, this expansion falls down when the modulation frequency $\omega$, or $l$th multiple of it, is comparable to one of the energy scale of the problem, say, the Hubbard interaction strength $U$. That is to say, $l\hbar\omega\approx U$, and we call it the $l$th resonance. Note that in this case, because $U-l\hbar\omega$ is a small energy scale, we should apply another unitary transformation 
\begin{align}\label{eq:R}
	\hat{R}(t) = \exp(i\sum_{j}l\omega t\hat{n}_{j\uparrow}\hat{n}_{j\downarrow}), 
\end{align}
which alters the interaction strength to an effective one $\tilde{U}=U-l\hbar\omega$. Moreover, since $\hat{R}(t)$ does not commute with the hopping term, it introduces an additional density dependence to the hopping term, and effectively it changes the gauge field to a spin and density dependent one as 
 \begin{align}
\tilde{\mathbf{A}}_{ij,\sigma}(t)&=\mathbf{A}(t)-\frac{l\omega t}{d^2} \mathbf{d_{ij}}((1-\hat n_{i\bar\sigma})\hat n_{j\bar\sigma}-(1-\hat n_{j\bar\sigma})\hat n_{i\bar\sigma})).
 \end{align}
Now the high frequency expansion can be safely applied, and to the lowest order it again results in a time-independent effective Hamiltonian written as
\begin{align} 
	\hat{H}_{\text{eff}} = \sum_{\langle i,j\rangle, \sigma} -\hat{J}^{\langle ij\rangle}_{\text{eff},\sigma}\hat{c}_{i\sigma}^{\dagger}\hat{c}_{j\sigma} + \tilde{U}\sum_{i}\left(\hat{n}_{i\uparrow}-\frac{1}{2}\right)\left(\hat{n}_{i\downarrow}-\frac{1}{2}\right).  \label{Heff}
\end{align}
Here $\hat{J}^{\langle ij\rangle}_{\text{eff},\sigma}$ is defined as 
\begin{equation} 
\hat{J}^{\langle ij\rangle}_{\text{eff},\sigma}=J_0\hat{a}_{ij\bar{\sigma}}+J_1\hat{b}_{ij\bar{\sigma}},
\end{equation}
where $\bar{\sigma}$ denotes the complement of $\sigma$, $J_0 = J\mathcal{B}_0(\mathcal{A})$, $J_1 = J \mathcal{B}_l(\eta_{ij}\mathcal{A})$ ($\eta_{ij}=\pm 1$ for $(i_x,i_y)=(j_x\pm1,j_y) \   \ \text{or} \   \ (i_x,i_y=j_x,j_y\pm 1)$), and
\begin{align}
\hat{a}_{ij\sigma} &= (1-\hat{n}_{i\sigma})(1-\hat{n}_{j\sigma}) + \hat{n}_{i\sigma}\hat{n}_{j\sigma},\\
	\hat{b}_{ij\sigma} &= (-1)^l(1-\hat{n}_{i\sigma})\hat{n}_{j\sigma} + \hat{n}_{i\sigma}(1-\hat{n}_{j\sigma}).
\end{align}
In above, the site dependence of $J_1$ is made implicitly. Note, however, that for even $l$ the Bessel function $\mathcal{B}_l$ is an even function, in which case $\eta_{ij}$ can be simply dropped and $J_1$ becomes a constant. Compared to the off-resonance case, now the hopping strength depends on site occupation of fermions. As we will show below, this correlated hopping plays a key role in the emergent new mechanism for ferromagnetism phase. 


\textit{Symmetry.} Before we discuss how to solve this effective Hamiltonian, let us first comment on the symmetry of this problem. 
Note that the original Hubbard model possesses a SO(4) symmetry \cite{so4}, which is composed of a spin SU(2), generated by $\hat{S}_z = (1/2)\sum_{i}\hat{c}_{i\uparrow}^{\dagger}\hat{c}_{i\uparrow}-\hat{c}_{i\downarrow}\hat{c}_{i\downarrow}$, $\hat{S}_{+} = \sum_{i}\hat{c}_{i\uparrow}^{\dagger}\hat{c}_{i\downarrow}$ and $\hat{S}_{-} = \hat{S}_{+}^{\dagger}$, and a charge SU(2), generated by $\hat{L}_z= -(1/2)\sum_{i}\hat{c}_{i\uparrow}^{\dagger}\hat{c}_{i\uparrow}+\hat{c}_{i\downarrow}^{\dagger}\hat{c}_{i\downarrow}+N_\text{s}/2$, $\hat{L}_{+}=\sum_{i}(-1)^i\hat{c}_{i\uparrow}\hat{c}_{i\downarrow}$ and $\hat{L}_{-} = \hat{L}_{+}^{\dagger}$. $N_\text{s}$ is the total number of sites. The spin SU(2) ensures that the direction of spin-density-wave (SDW) order parameter can be taken along any direction, while the charge SU(2) ensures the degeneracy of a charge-density-wave (CDW) order and the fermion pairing order (P). 

In the presence of periodic modulation, considering the time-dependent Hamiltonian Eq. \ref{eq:H}, it is straightforward to show that the spin SU(2) symmetry stays, yet the charge SU(2) symmetry no longer holds because $\hat{L}_z$ does not commute with the $\sum_{i,\sigma}f_{i}(t)\hat{n}_{i\sigma}$ term. However, considering the time-independent effective Hamiltonian Eq. \ref{Heff}, one can show that the charge SU(2) symmetry is recovered for even $l$ case though not for odd $l$ case \cite{supple}. Hereafter we focus only on the even $l$ case which possesses the same SO(4) symmetry as the original Hubbard model. In addition, the effective Hamiltonian also possesses particle-hole symmetry at half-filling. 

\begin{figure}[t]
	\centering
	\includegraphics[width=1.\columnwidth]{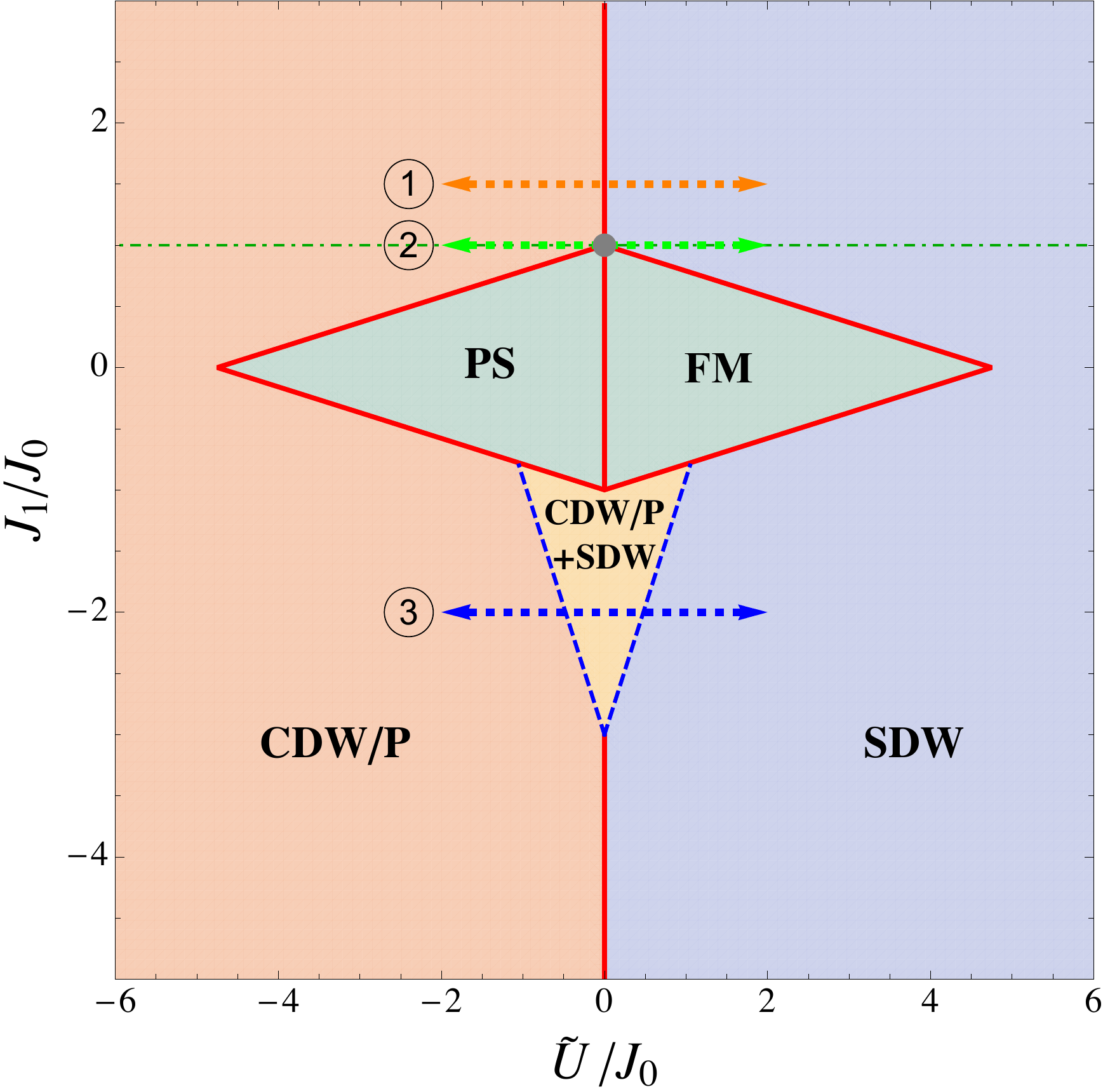}
	\caption{Phase diagram for the effective Hamiltonian Eq. \ref{Heff} of a resonantly driven Fermi Hubbard model with even $l$ at half filling. The phase diagram is controlled by two dimensionless parameters, $J_1/J_0$ and $\tilde{U}/J_0$. ``CDW" and ``SDW" denote charge and spin density wave order. ``P" denote fermion pairing order. ``FM" denotes ferromagnetism, ``PS" denotes phase separation into high and low density regimes. Red lines denote the first order transition and the blue dashes lines denote the second order transition. Dashed arrows mark the lines along which the order parameters are plotted in FIG. \ref{fig:order_parameter}. }\label{fig:phasediagram}
\end{figure}

\textit{Phase Diagram.} We present our results on the phase diagram following from a standard mean-field treatment based on this effective Hamiltonian Eq. \ref{Heff}, which is known to be qualitatively reliable for a normal Hubbard model \cite{supple,nagaosa}. Thanks to the SO(4) symmetry, we can choose SDW along $\hat{z}$ direction (i.e. $s_i=\langle \hat{n}_{i\uparrow}-\hat{n}_{i\downarrow}\rangle$) and CDW (i.e. $c_i=\langle \hat{n}_{i\uparrow}+\hat{n}_{i\downarrow}\rangle-1$) as the order parameters in our mean-field theory.  Note that when we obtain the CDW order, it means that the system can have either CDW order or fermion pairing order, or an arbitrary combination of them, as the order parameter of the degenerate ground states. Higher order effect will break the degeneracy between CDW and fermion pairing order, but it is beyond the scope of current work. 

\begin{figure}[t]
	\centering
	\includegraphics[width=0.9\columnwidth]{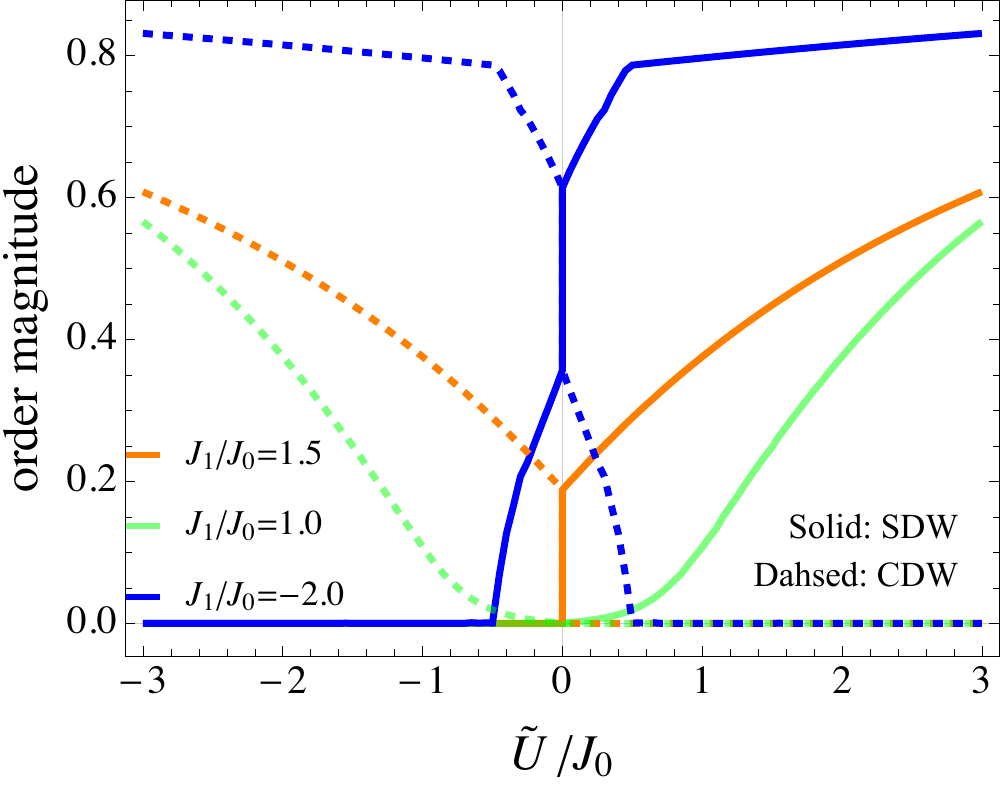}
	\caption{The CDW order parameter $c$ (dashed lines) and the SDW order parameter $s$ (solid lines) as a function of $\tilde{U}$ for three representative cases labeled as 1-3 in Fig. \ref{fig:phasediagram}, with $J_1/J_0=1.5$, $1.0$ and $-2.0$, respectively.    }\label{fig:order_parameter}
\end{figure}

The phase diagram is shown in Fig. \ref{fig:phasediagram}. Setting $J_0$ as the energy unit, the phase diagram is controlled by two parameters of $J_1/J_0$ and $\tilde{U}/J_0$, both of which can be easily tuned from positive to negative via control of $\omega$ and $A$. As breach-mark of our calculation, first of all, note that when $J_1/J_0=1$, because $\hat{a}_{ij\sigma}+\hat{b}_{ij\sigma}=\hat{I}$, the kinetic energy term becomes $-J_0\sum_{\langle ij\rangle,\sigma}\hat{c}^\dag_{i\sigma}\hat{c}_{j\sigma}$ and the Hamiltonian recovers the usual Hubbard model. In this case (labeled by 2 in Fig. \ref{fig:phasediagram}), the result for the normal Hubbard model is retrieved where we obtain a CDW order of $(\pi,\pi)$ with attractive interaction ($\tilde{U}<0$) and a SDW order of $(\pi,\pi)$ with repulsive interaction ($\tilde{U}>0$). Explicitly, the order parameters are chosen as $s_i=(-1)^{i_x+i_y}s$ and $c_i=(-1)^{i_x+i_y}c$, and $s$($c$) gradually vanishes as $\tilde{U}$ approaches zero from the positive (negative) side. As a result, a second order phase transition occur at $\tilde{U}=0$ (gray dot in FIG. \ref{fig:phasediagram}). This can be seen from the order parameters plotted as a function of $\tilde{U}$, shown as green curves in Fig. \ref{fig:order_parameter}.

Since both CDW and SDW are ordered phases, the more generic situation should be either a first order transition or a phase co-existence regime in between. As marked by the red solid lines in Fig. \ref{fig:phasediagram}, the phase boundary of CDW and SDW at large $|J_1/J_0|$ is a first order transition. The order parameters are shown with orange lines in Fig. \ref{fig:order_parameter} for a representative case (labeled by 1 in Fig. \ref{fig:phasediagram}), where the CDW or SDW order parameter jumps from a finite value to zero at $\tilde{U}=0$. At certain regime of $J_1/J_0$, a CDW and SDW co-existence regime shows up in between as displayed by the yellow regime in Fig. \ref{fig:phasediagram}. The order parameters are shown with blue lines in Fig. \ref{fig:order_parameter} for a representative case (labeled by 3 in Fig. \ref{fig:phasediagram}).

\begin{figure}[t]
	\centering
	\includegraphics[width=0.9\columnwidth]{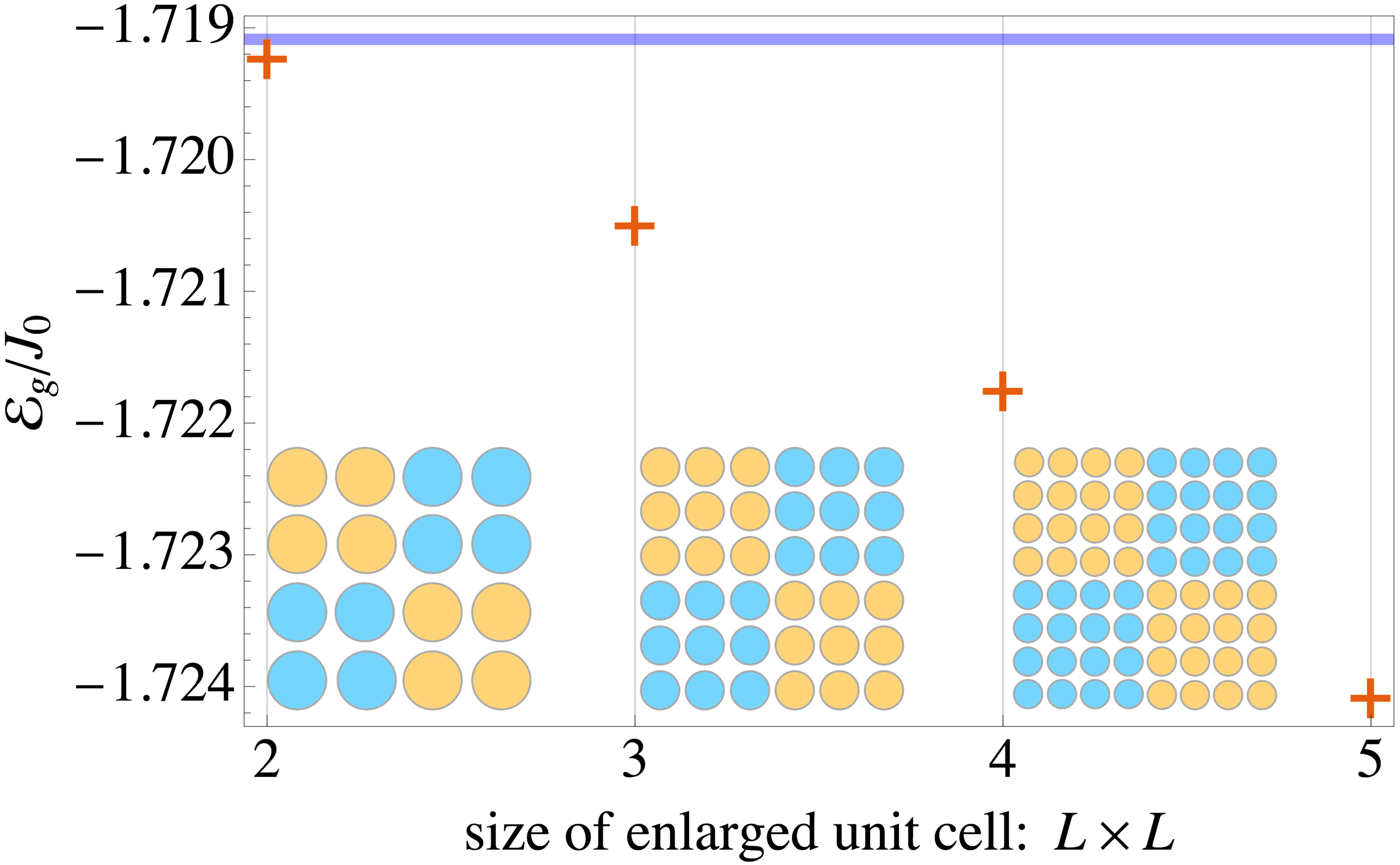}
	\caption{The mean-field ground state energy as a function of the domain size $L$. The insets show the configuration for $2\times 2$, $3\times 3$ and $4\times4$ blocks where the CDW and SDW order parameters are uniformly distributed within a domain and take opposite values between neighboring domains. For comparison, the solid line is the mean-field energy when order parameters are all zero.   }\label{fig:phase_seperation}
\end{figure}

The most notable feature in Fig. \ref{fig:phasediagram} is the green region. In this region a mean-field ansatz of CDW or SDW orders with ordering vector at $(\pi,\pi)$ may not yield any ordered solution. However, when we consider the case of enlarged $2\times 2$, $3\times 3$, up to $L\times L$ domains, and within each domain the CDW and SDW order parameters are uniformly chosen as $c$ and $s$ while in its neighboring domain they are taken as $-c$ and $-s$, the mean-field ansatz does yield ordered solutions. It can be seen from Fig. \ref{fig:phase_seperation} which shows that the mean-field ground state energy decreases monotonically as $L$ increases. It indicates that the ground state will form large domains with opposite order parameter values. Moreover, minimizing ground state energy yields $c=0$ and $s\neq 0$ at positive $\tilde{U}$ and $c\neq 0$ and $s=0$ at negative $\tilde{U}$. Hence, the system at positive $\tilde{U}$ possesses spin order, and the increasing of the domain size means the decreasing of the spin ordering wave vector. Eventually, the wave vector decreases toward zero, and the ground state becomes a ferromagnetic state. In another word, as the domain size becomes larger and larger, the system is essentially made of ferromagnetic domains. For negative $\tilde{U}$ the system tends to phase separation with high density in one domain and low density in its neighboring domain. The transition between the ferromagnetic phase to the SDW phase, as well as the transition from phase separation to CDW, is a first order transition.   

It should be emphasized that both the ferromagnetism and the phase separation regime occur at small $\tilde{U}$. In fact, it is purely due to the correlated hopping effect in the effective Hamiltonian Eq. \ref{Heff}, which originates essentially from the resonant driving. Considering the mean-field configurations as shown in the insets of Fig. \ref{fig:phase_seperation}, let us look at the mean-field value of the effective hopping strength $\langle \hat{J}^{\langle ij\rangle}_{\text{eff},\sigma}\rangle $ which quantifies how the particle occupations affect the hopping strength and thus affect the bandwidth. We define $J^{\text{intra}}_{\text{eff},\sigma}$ as $\langle \hat{J}_{\text{eff},\sigma}^{\langle ij\rangle}\rangle$ with both $i$ and $j$ in the same domain, and $J^{\text{inter}}_{\text{eff},\sigma}$ as $\langle \hat{J}_{\text{eff},\sigma}^{\langle ij\rangle}\rangle$ with $i$ and $j$ across two neighboring domains. It is straightforward to write down both  $J^{\text{intra}}_{\text{eff},\sigma}$ and $J^{\text{inter}}_{\text{eff},\sigma}$ as 
\begin{align}
	J^{\text{intra}}_{\text{eff},\sigma} &= \frac{1}{2} \left(J_0[1+(c\mp s)^2]+J_1[1-(c\mp s)^2]\right), \\
	J^{\text{inter}}_{\text{eff},\sigma} &= \frac{1}{2} \left(J_0[1-(c\mp s)^2]+J_1[1+(c\mp s)^2]\right),
\end{align}
where $\mp$ corresponds to different spin component. One can show that when $|J_1|< |J_0|$, $|J^{\text{inter}}_{\text{eff},\sigma}| $ is always smaller than $|J^{\text{intra}}_{\text{eff},\sigma}|$. Hence, the size of the domain tends to increase such that there are more intra-domain links than inter-domain links, and therefore the effective bandwidth on average becomes larger. For a given filling, a larger bandwidth leads to more kinetic energy gain. $|J_1|< |J_0|$ is precisely the regime where we find ferromagnetism or phase separation in the phase diagram of Fig. \ref{fig:order_parameter} with arbitary weak interaction. This regime can be easily accessed when $\mathcal{A}$ is small. 

\begin{figure}[t]
	\centering
	\includegraphics[width=1.\columnwidth]{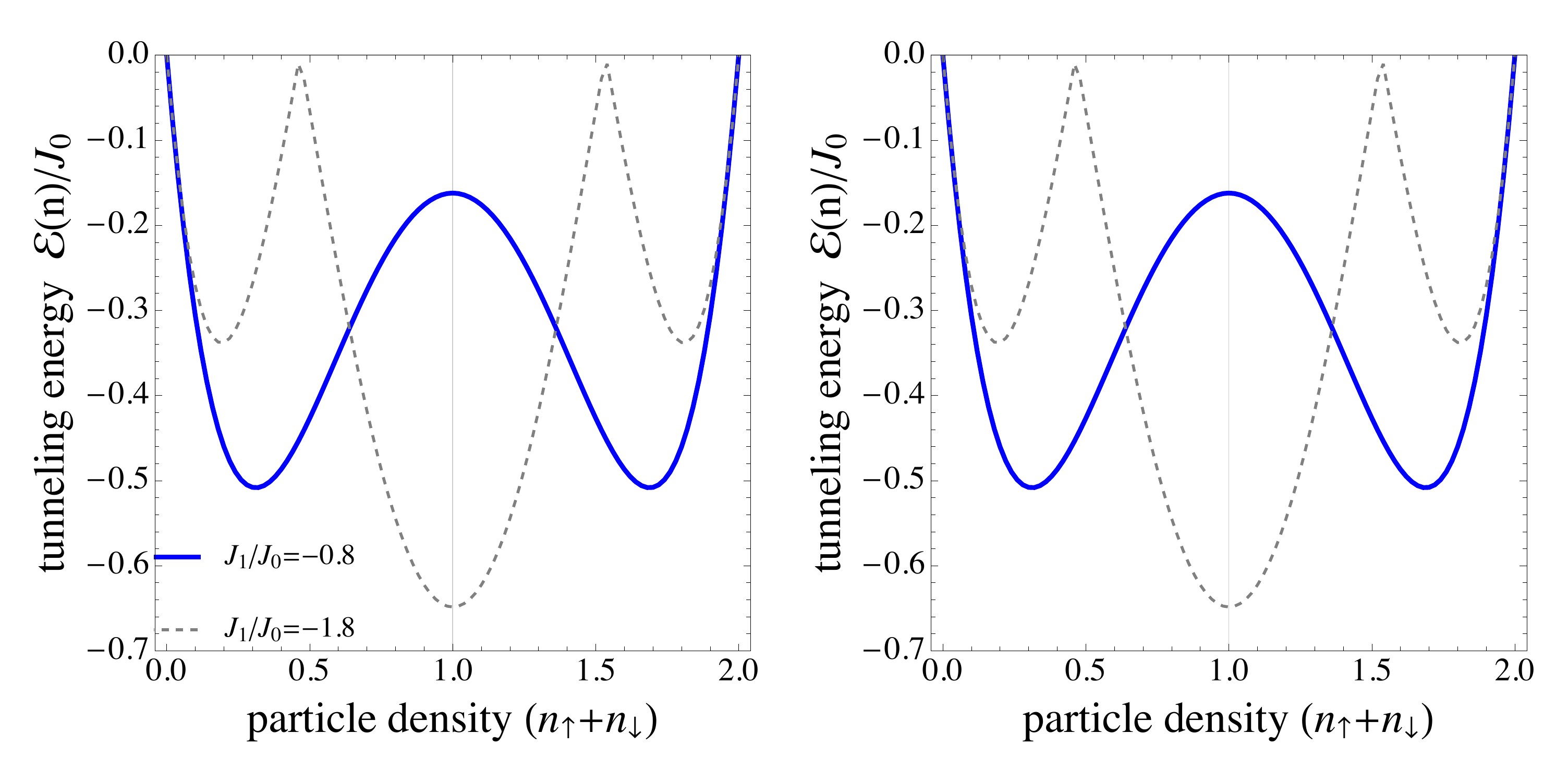}
	\caption{The mean-field energy with $\tilde{U}=0$ (a) as a function of total density $n$ with $s_z=0$ fixed, and (b) as a function of $s_z$ with $n=1$ fixed. Solid line is for $J_1/J_0=-0.8$ and the dashed line is for $J_1/J_0=-1.8$. }\label{fig:kinetic_energy}
\end{figure}

An alternative way to understand the emergence of this ferromagnetism is to consider a uniform system with $\tilde{U}=0$, where the Hamiltonian contains only the correlated tunneling term. Substitute $ \hat{J}^{\langle ij\rangle}_{\text{eff},\sigma} $ by its mean-field value, it is straightforward to compute the kinetic energy of this uniform system that depends on $n_{i\uparrow}$ and $n_{i\downarrow}$, where $n_{\uparrow}=(n+s_z)/2$ and $n_{\downarrow}=(n-s_z)/2$. We plot the kinetic energy as a function of $n$ for $s_z=0$ in Fig. \ref{fig:kinetic_energy}(a) and as a function of $s_z$ for $n=1$ in Fig. \ref{fig:kinetic_energy}(b) for two representative cases with $J_1/J_0=-0.8$ and $-1.8$. One can see from Fig. \ref{fig:kinetic_energy}(a) that for $J_1/J_0=-0.8$, there are two local minima with one at $n>1$ and the other at $n<1$, who locate symmetrically on two sides of $n=1$, while for $J_1/J_0=-1.8$ there is only one minimum located at $n=1$. Similarly, in Fig. \ref{fig:kinetic_energy}(b) for $J_1/J_0=-0.8$, there are two local minima with one at positive $s_z$ and the other at negative $s_z$, symmetrically distributed around $s_z=0$, and for $J_1/J_0=-1.8$ there is only one minimum at $s_z=0$. Thus, when the system is constrained with the average $n=1$ and $s_z=0$, for the case with $J_1/J_0=-0.8$, it will actually phase separate into domains with either different $n$ or different $s_z$, corresponding to phase separation and ferromagnetism, respectively. The choice is made by the sign of $\tilde{U}$ when a small but finite $\tilde{U}$ perturbation is turned on.

\textit{Conclusion and Outlook.} The most significant finding of this work is to provide an alternative mechanism for the onset of ferromagnetism in the model with correlated hopping, which roots in the cooperation between the spin order and the correlated hopping. It is driven by the kinetic energy term, and occurs in the weakly interacting regime. This is in contrast to the well-known Stoner ferromagnetism mechanism which is driven by large interaction energy and occurs when the interaction strength is beyond certain critical value. This is also different from the ferromagnetism due to the super-exchange processes discussed in the experiment of Ref. \cite{ETH_driven} which requires $\tilde{U}$ to be negative. 

Finally we shall comment on the experimental observation of this ferromagnetism. First of all, the system itself has been realized with cold atoms, and moreover, a very recent experiment shows that the heating is insignificant in the presence of driving and the life time of the system can be about one second  \cite{new_exp}. Second, because this ferromagnetism is driven by the kinetic energy, and as one can see from Fig. \ref{fig:kinetic_energy}, the energy gain is of the order of bandwidth, thus one expects this ferromagnetism to be observed when temperature is of the order of bandwidth, which can be accessed now by cold atom experiments. Thirdly, the quantum gas microscope techniques mentioned at the beginning can be used to detect real space ferromagnetic domains. Hence, it is quite promising to verify this theory experimentally in very near future. 

\textit{Acknowledgment.} This work is supported MOST under Grant No. 2016YFA0301600 and NSFC Grant No. 11734010.

\end{document}


\title{Supplementary Material: Resonant Driving induced Ferromagnetism in the Fermi Hubbard Model}

\author{Ning Sun}
\affiliation{Institute for Advanced Study, Tsinghua University, Beijing, 100084, China}
\author{Pengfei Zhang}
\affiliation{Institute for Advanced Study, Tsinghua University, Beijing, 100084, China}
\author{Hui Zhai}
\affiliation{Institute for Advanced Study, Tsinghua University, Beijing, 100084, China}
\affiliation{Collaborative Innovation Center of Quantum Matter, Beijing, 100084, China}

\date{\today }
\begin{abstract}

In this supplementary material, we first discuss the symmetry of the resonantly driven Fermi Hubbard model. The SO(4) symmetry \cite{so4} of the effective Hamiltonian for even $l$ is proved. Then we establish the mean-field theory adopted in this work for even $l$, which is consistent with the text-book results \cite{Altland} for $J_0=J_1$.
\end{abstract}
\maketitle

\section{Symmetry}
In this section we dicuss the symmetry of the resonantly driven Fermi Hubbard model. The symmetry is of vital significance since it greatly simplifies the formulation of the mean-field description. 
\paragraph{SO(4) symmetry.---}
A usual bipartite Fermi-Hubbard model, written as $\hat{H}=-J\sum_{\langle i,j\rangle,\sigma}\hat{c}_{i\sigma}^{\dagger}\hat{c}_{j\sigma}+\tilde{U}\sum_{i}(\hat{n}_{i\uparrow}-1/2)(\hat{n}_{i\downarrow}-1/2)$, possesses SO(4) symmetry \cite{so4}. The SO(4) symmetry is resolved into two SU(2) symmetries, the spin SU(2) and the charge SU(2). The generators of spin SU(2) are given by: 
\begin{align}\label{eq:spinsu2}
	\hat{S}_z = \frac{1}{2}\sum_{i}\hat{c}_{i\uparrow}^{\dagger}\hat{c}_{i\uparrow}-\hat{c}_{i\downarrow}\hat{c}_{i\downarrow}, \  \  
	\hat{S}_{+} = \sum_{i}\hat{c}_{i\uparrow}^{\dagger}\hat{c}_{i\downarrow}, 
\end{align}
and
\begin{align}
	\hat{S}_{-} = \hat{S}_{+}^{\dagger}, \  \  
	\hat{S}_{x} = \frac{\hat{S}_{+}+\hat{S}_{-}}{2}, \  \  
	\hat{S}_{y} = \frac{\hat{S}_{+}-\hat{S}_{-}}{2i},
\end{align}
who satisfy the commutation relation of the SU(2) algebra. Due to the contraction between spin indices, the very beginning time-dependent Hamiltonian $\hat{H}(t)$ (Eq.(4) in the main text) is invariant under this SU(2) symmetry operations, is also the unitary transformations $\hat{R}(t)$ (Eq.(5) and (6) in the main text). Since the time average does not alter this attribute, one arrives at the conclusion that the effective static Hamiltonian $\hat{H}_{\text{eff}}$ in Eq.(7) of the main text also possess this spin SU(2) symmetry no matter $l$ even or odd. This can also be verified directly by checking $\left[\hat{H}_{\text{eff}},\hat{S}_{\alpha}\right]=0$ for any $\alpha=x,y,z$. The spin rotational symmetry hence allows us to automatically get a spin-balanced system without any additional magnetic field.

However, regarding the charge SU(2) symmetry, the time-dependent Hamiltonian lacks it for the appearance of the time-dependent onsite energy term (the last term in Eq.(4) of the main text). As a result we do not expect the SO(4) symmetry in general. Nevertheless, it emerges in the effective Hamiltonian with $l$ even. We introduce the charge SU(2) as follows \cite{so4}. 
\begin{align}
	\hat{L}_z &= -\frac{1}{2}\sum_{i}\hat{c}_{i\uparrow}^{\dagger}\hat{c}_{i\uparrow}+\hat{c}_{i\downarrow}^{\dagger}\hat{c}_{i\downarrow}+\frac{1}{2}N, \\ 
	\hat{L}_{+} &= \sum_{i}(-1)^i\hat{c}_{i\uparrow}\hat{c}_{i\downarrow} = \sum_{i}\exp(i\mathbf{Q}\cdot\mathbf{x}_i)\hat{c}_{i\uparrow}\hat{c}_{i\downarrow}
\end{align}
where $\mathbf{Q} = (\pi, \pi)$, $N$ is the total number of lattice sites hereinafter, 
and
\begin{align}
	\hat{L}_{-} = \hat{L}_{+}^{\dagger}, \  \  
	\hat{L}_{x} = \frac{\hat{L}_{+}+\hat{L}_{-}}{2}, \  \  
	\hat{L}_{y} = \frac{\hat{L}_{-}-\hat{L}_{-}}{2i}. 
\end{align}
$\{\hat{L}_{z}, \hat{L}_{x}, \hat{L}_{y}\}$ forms an SU(2) algebra. It can be verified straightforwardly that the effective Hamiltonian $\hat{H}_{\text{eff}}$ with $l$ even commutes with all these generators. This fact can also be seen from the following transformation 
$$\hat P: \hat c_{i\uparrow}\rightarrow (-1)^i\hat c^\dagger_{i\uparrow},$$ who maps $\hat{H}_{\text{eff}}$ 
with $l$ even of interaction $+\tilde{U}$ to the same class but of effective interaction $-\tilde{U}$, exchanging the role played by $\{L_i\}$ and $\{S_i\}$. As a result, the spin SU(2) invariance for a model with $-\tilde{U}$ indicates the charge SU(2) symmetry for $+\tilde{U}$. And vice versa. In addition, this transformation also implies the phase diagram should be symmetric under $+\tilde{U} \leftrightarrow -\tilde{U}$ together with the interchange of charge and spin order. However, in the $l$ odd cases, this mapping falls down because of the additional minus sign in $\hat{b}_{ij\sigma}$. 

\paragraph{Particle-hole symmetry.---} The half-filled effective Hamiltonian $\hat{H}_{\text{eff}}$ holds particle-hole symmetry no matter $l$ even or odd. (i) When $l$ is even, define particle-hole transformation $\hat{C}: \hat{c}_{i\sigma} \rightarrow (-1)^i\hat{c}_{i\sigma}^{\dagger}$. The Hamiltonian is invariant under $\hat{C}$. $[\hat{C}, \hat{H}_{\text{eff}}]=0$. (ii) When $l$ is odd, we further define the bipartite transformation $\hat{S}$, which switches the $A/B$ sublattices. Then the Hamiltonian is invariant under the combination of $\hat{C}$ and $\hat{S}$: $[\hat{C}\hat{S}, \hat{H}_{\text{eff}}]=0$. This symmetry allow us to fix chemical potential $\mu=0$ during the mean-field calculation at half-filling situation.

\section{Mean-field treatment}
In this section we establish the mean-field theory adopted in this work. As explained in the main text, assuming no canted order, we can consider only the charge density wave (CDW) order and spin density wave (SDW) order in the $z$ direction because of the SO(4) symmetry. We focus on the half-filled spin-balanced system, in which case the total charge density and total magnetic momentum is given by  
\begin{align}
	\langle \hat{n} \rangle = 1, \  \  \ \ \ 
	\langle \hat{S} \rangle = 0. 
\end{align}

\paragraph{Path integral approach.---} Here we provide a derivation of the mean-field Hamiltonian Eq. \eqref{eq:Hmf} via a path integral approach. In the path integral language, the real-time partition function of the system is given by:
\begin{align}
\mathcal{Z}=&\int \mathcal D\psi D\bar\psi \exp(i\int dt L)  \\
L=&\sum_{i,\sigma} i\bar \psi_{i\sigma}\partial_t\psi_{i\sigma}+\sum_{\langle i,j\rangle, \sigma} \left(J_0a_{ij\bar{\sigma}}(\bar\psi \psi) + J_1b_{ij\bar{\sigma}}^l(\bar\psi\psi)\right)\bar \psi_{i\sigma}\psi_{j\sigma} \notag \\&- \tilde{U}\sum_{i}\bar\psi_{i\uparrow}\psi_{i\uparrow}\bar\psi_{i\downarrow}\psi_{i\downarrow} \  .
\end{align}
Here $\psi$ corresponds to the fermionic field operators, and $a_{ij\bar{\sigma}}$ and $b_{ij\bar{\sigma}}^l$ are given by replacing the operators in $\hat a_{ij\bar{\sigma}}$ and $\hat b_{ij\bar{\sigma}}^l$ with fields. Since the effective Hamiltonian (Eq. (7) in the main text) contains six-fermion terms, the traditional decoupling based on Hubbard-Stratonovich transformation does not applied directly. However, an alternative way of decoupling based on the similar spirit of Hubbard-Stratonovich transformation should be adopted here. Specifically, by inserting a delta function, we directly introduce the auxiliary bosonic field:  
\begin{align}
\mathcal{Z}=&\int \mathcal D\psi D\bar\psi Dn \prod_{i\sigma}\delta(n_{i,\sigma}-\bar\psi_{i,\sigma}\psi_{i,\sigma})\exp(i\int dt L)\\
L=&\sum_{i,\sigma} i\bar \psi_{i\sigma}\partial_t\psi_{i\sigma}+\sum_{\langle i,j\rangle, \sigma} \left(J_0a_{ij\bar{\sigma}}(n) + J_1b_{ij\bar{\sigma}}^l(n)\right)\bar \psi_{i\sigma}\psi_{j\sigma}  \notag \\&- \tilde{U}\sum_{i}n_{i\uparrow}n_{i\downarrow} \  .
\end{align}
Thanks to the delta function inserted, one could replace all $\psi^\dagger \psi$ by $n$ in the action. As an equivalence check, if we integrate out $n$ fields first, we get our original action back. Now we introduce another field $\eta$ to absorb the delta function into an integral:
\begin{align}
\mathcal{Z}=&\int \mathcal D\psi D\bar\psi Dn D\eta\exp(i\int dt L)\\
L=&\sum_{i,\sigma} i\bar \psi_{i\sigma}\partial_t\psi_{i\sigma}+\sum_{\langle i,j\rangle, \sigma} \left(J_0a_{ij\bar{\sigma}}(n) + J_1b_{ij\bar{\sigma}}^l(n)\right)\bar \psi_{i\sigma}\psi_{j\sigma}  \notag \\&- \tilde{U}\sum_{i}n_{i\uparrow}n_{i\downarrow}-\sum_{i \sigma}\eta_{i\sigma}(n_{i,\sigma}-\bar\psi_{i,\sigma}\psi_{i,\sigma})
\label{eq:pathint_-1}
\end{align}
As a result, the fermion becomes quadratic. In general, one can integrate out all the fermions to get an effective action of wholly bosonic degrees of freedom. Whereas, the mean-field approximation is to say that all the bosonic fields will be replaced by their saddle point solutions.

\paragraph{Mean-field Hamiltonian.---} 
By doing a Legendre transformation and replacing the fermion bilinears with its expectation values, we obtain the mean-field Hamiltonian
\begin{align}\label{eq:Hmf0}
	\hat{H} = \sum_{\langle i,j\rangle, \sigma} - \left(J_0a_{ij\bar{\sigma}}(n) + J_1b_{ij\bar{\sigma}}^l(n)\right)\hat{\psi}^{\dagger}_{i\sigma}\hat{\psi}_{j\sigma}  + \tilde{U}\sum_{i}n_{i\uparrow}n_{i\downarrow}+\sum_{i \sigma}\eta_{i\sigma}(n_{i,\sigma}-\hat{\psi}^{\dagger}_{i,\sigma}\hat{\psi}_{i,\sigma}) 
\end{align}
where the following replacement is carried out:
 \begin{align}
	\hat{a}_{ij\sigma} &= (1-\hat{n}_{i\sigma})(1-\hat{n}_{j\sigma}) + \hat{n}_{i\sigma}\hat{n}_{j\sigma}, \rightarrow (1-n_{i\sigma})(1-n_{j\sigma}) + n_{i\sigma}n_{j\sigma}\\
	\hat{b}_{ij\sigma} &= (-1)^l(1-\hat{n}_{i\sigma})\hat{n}_{j\sigma} + \hat{n}_{i\sigma}(1-\hat{n}_{j\sigma})\rightarrow (-1)^l(1-n_{i\sigma})n_{j\sigma} + n_{i\sigma}(1-n_{j\sigma})
 \end{align}

For the half-filled spin-balanced situation, we focus on the following two cases here. Notice that, in a half-filled spin-balanced system, the local particle density can be expressed in terms of local charge density and local spin density as
\begin{align}
	n_{i\uparrow} = \frac{1}{2}(1 + c_i + s_i ) \;,\;\; n_{i\downarrow} = \frac{1}{2}(1+c_i-s_i)
\end{align}
where $\sum_ic_i=0$ and $\sum_is_i=0$ . 
The two cases are then: 
\begin{enumerate}
	\item $c_i=(-1)^{i_x+i_y}c$, $s_i=(-1)^{i_x+i_y}s$
	\item $c_i=(-1)^{\lceil i_x/L \rceil + \lceil i_y/L \rceil}c, s_i=(-1)^{\lceil i_x/L \rceil + \lceil i_y/L \rceil}s$  
\end{enumerate}
where $c$ and $s$ are CDW and SDW order parameters introduced, and $\lceil x \rceil$ denotes the ceiling function of $x$. 
Substituting these two expressions into the original mean-field Hamiltonian Eq.(\ref{eq:Hmf0}) will yield the mean-field Hamiltonian in each case. 

\begin{figure}[t]
	\centering
	\includegraphics[width=.8\textwidth]{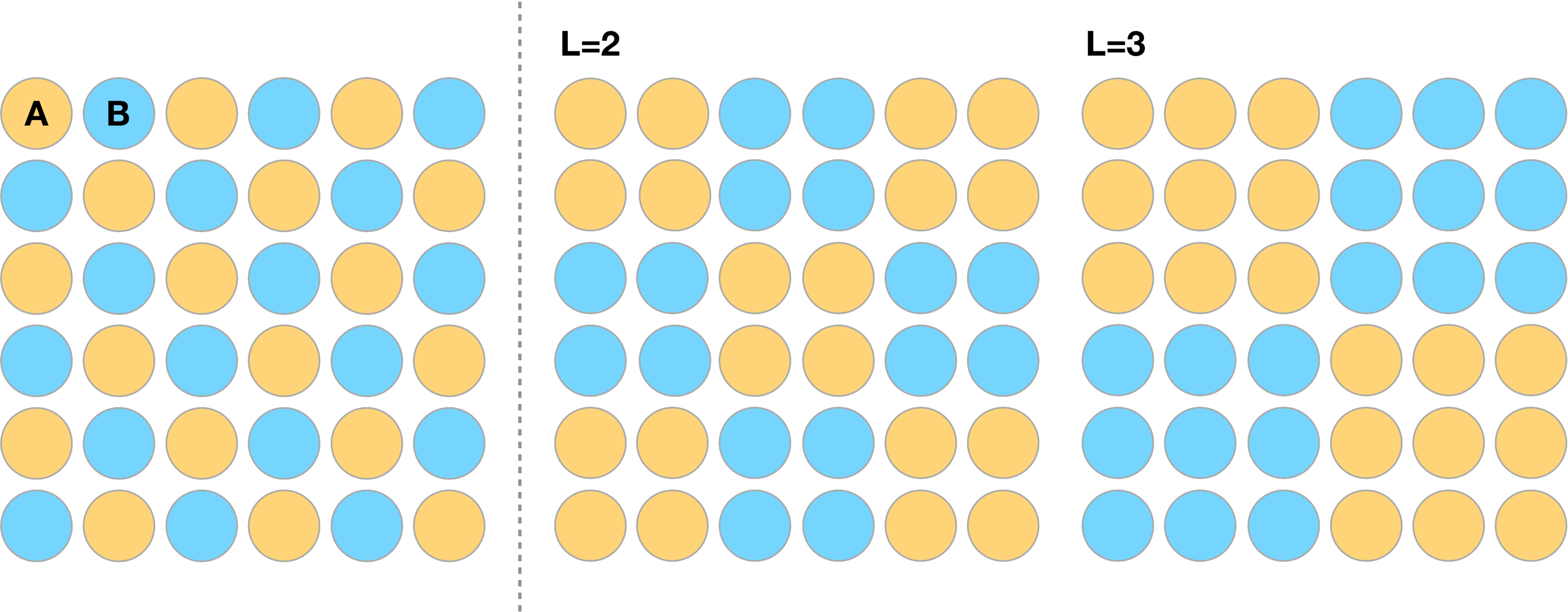}
	\caption{Schematic of the two cases. Left panel: case 1. Right panel: typical examples of case 2 when $L=2$ and $L=3$, respectively, as labeled. Lattice sites of the same color have the same local particle density $c_i$ and local spin density $s_i$. }\label{fig:twocases}
\end{figure}

\paragraph{Case 1.} The first case is actually the $(\pi,\pi)$ order, where the lattice is naturally divided into $A/B$ sublattices and the unit cell of the lattice consists of 2 sites, with one from $A$ and the other from $B$. The schematic is shown in FIG. \ref{fig:twocases} (Left). 

To be specific, we explicit show here the workflow in the first case. By substitution, 
\begin{align}
	&n_A\equiv\langle \hat{n}_{i\in A} \rangle = 1 + c , \  \  \
	s_A\equiv\langle \hat{S}_{i\in A}^{z} \rangle = s/2,\label{order1} \\
	&n_B\equiv\langle \hat{n}_{i\in B} \rangle = 1 - c , \  \  \
	s_B\equiv\langle \hat{S}_{i\in B}^{z} \rangle = -s/2,\label{order2}
\end{align}
and the mean-field Hamiltonian is explicitly written as: 
\begin{align}\label{eq:Hmf}
	\hat{H}_{\text{mf}} &= \sum_{\langle i,j\rangle,\sigma} \left\{-J_0[(1-n_{A\bar{\sigma}})(1-n_{B\bar{\sigma}})+n_{A\bar{\sigma}}n_{B\bar{\sigma}}] - J_1[(1-n_{A\bar{\sigma}})n_{B\bar{\sigma}}+n_{A\bar{\sigma}}(1-n_{B\bar{\sigma}})]\right\}\hat{c}_{i\sigma}^{A\dagger}\hat{c}_{j\sigma}^B+ \text{H. C. } + \frac{\tilde{U}N}{2}(n_{A\uparrow}n_{A\downarrow} + n_{B\uparrow}n_{B\downarrow}) \notag \\
		& \  \  + \eta_c \left[c-\frac{(\hat{n}_{A\uparrow}+\hat{n}_{A\downarrow}) - (\hat{n}_{B\uparrow}+\hat{n}_{B\downarrow})}{2}\right] 
		+ \eta_s \left[s-\frac{(\hat{n}_{A\uparrow}-\hat{n}_{A\downarrow}) - (\hat{n}_{B\uparrow}-\hat{n}_{B\downarrow})}{2}\right]
\end{align}
where $\hat{c}_{i\sigma}^A$($\hat{c}_{i\sigma}^B$) are the annihilation operators of $i$ site and spin $\sigma$ on $A$($B$) sublattice, $\hat{n}_{\mu\sigma} = \frac{1}{N/2}\sum_{i\in\mu}\hat{n}_{i\sigma}$, $\mu = A$ or $B$, and $n_{\mu\sigma} = \langle\hat{n}_{\mu\sigma}\rangle$. Here $\eta_c$ and $\eta_s$ are Lagrangian multipliers of the CDW and SDW orders, respectively, who are also variational parameters to optimize the ground-state energy of $\hat{H}_{\text{mf}}$ to get a mean-field solution. 

After some substitution and simplification, the mean-filed Hamiltonian in momentum space is written as
\begin{align}\label{eq:Hmfk}
	\hat{H}_{\text{mf}} &= \sum_{\mathbf{k}\sigma}-P_{\sigma}(c,s)Q(\mathbf{k})\hat{c}_{\mathbf{k}\sigma}^{A\dagger}\hat{c}_{\mathbf{k}\sigma}^B+ \text{H.c.} 
		+ \frac{\tilde{U}N}{2}(1+c^2-s^2) \notag\\
		& \  \  \ 
		+ \eta_c \left[c-\frac{(\hat{n}_{A\uparrow}+\hat{n}_{A\downarrow}) - (\hat{n}_{B\uparrow}+\hat{n}_{B\downarrow})}{2}\right]
		+ \eta_s \left[s-\frac{(\hat{n}_{A\uparrow}-\hat{n}_{A\downarrow}) - (\hat{n}_{B\uparrow}-\hat{n}_{B\downarrow})}{2}\right]
\end{align}
where $\hat{c}_{\mathbf{k}\sigma}^A$($\hat{c}_{\mathbf{k}\sigma}^B$) are the annihilation operators on $A$($B$) sublattice of quasi-momentum $\mathbf{k}$ and spin $\sigma$, $Q(\mathbf{k}) = \sum_{i}\exp(i\mathbf{k}\cdot\mathbf{d}_i)$, $\{\mathbf{d}_i\}$ are the lattice vectors, and 
\begin{align}
	P_{\uparrow}(c,s) &= \frac{J_0}{2}\left[1-(c-s)^2\right] + \frac{J_1}{2}\left[1+(c-s)^2\right] \\
	P_{\downarrow}(c,s) &= \frac{J_0}{2}\left[1-(c+s)^2\right] + \frac{J_1}{2}\left[1+(c+s)^2\right] 
\end{align} 
The summation of $\mathbf{k}$ in Eq.(\ref{eq:Hmfk}) is over the first Brillioun zone. Minimization the energy for all parameters $\{c, s, \eta_c, \eta_s\}$ yields a set of mean-field equations that is solved by numerical iteration in this work.

\paragraph{Case 2.} 
The second case actually includes a series of circumstances of $L$ being $2, 3, 4, \ldots, \infty$. Typical examples of $L=2$ and $L=3$ are shown in the right panel of FIG. \ref{fig:twocases}. In this case, for each independent $L$, doing the similar substitution as above and solving optimization problem in a numerical way returns us a set of self-consistent mean-field solutions in each $L$. The ground state should be the one with the lowest ground state energy. 

FIG. 2, 3, 4, 5 in the main text are based on the mean-field numerics of above two cases. 

\paragraph{Check the formulation for $J_0=J_1$.---} 
Finally, we explain why our mean-field theory reduces to a traditional one appearing in common text books, e.g. \cite{Altland}. In fact, the $P_\sigma$ in this case is just a constant with no $c$ and $s$ dependence. Thus the variation of $c$ and $s$ yields:
\begin{align}
\eta_c=-\tilde{U}Nc\ \ \  \ \eta_s=-\tilde{U}Ns.
\end{align}

Using this relation, we see the mean-field Hamiltonian reduces to a familiar one that appears in textbooks \cite{Altland}.